\documentclass[10pt,letterpaper]{article}

\usepackage[squaren]{SIunits}
\usepackage{multirow}
\usepackage[utf8]{inputenc}
\usepackage{amsmath,amssymb}
\usepackage{cite}
\usepackage{nameref,url}
\usepackage{microtype}

\usepackage{rotating}
\usepackage{booktabs}

\date{}

\begin{document}
\vspace*{0.35in}

\begin{flushleft}
{\Large
\textbf\newline{Classification and Verification of Online Handwritten Signatures with Time Causal Information Theory Quantifiers}
}
\newline
\\
Osvaldo A.\ Rosso\textsuperscript{1,*},
Raydonal Ospina\textsuperscript{2},
Alejandro C.\ Frery\textsuperscript{3},
\\
\bigskip
\bf{1} Instituto de F\'{\i}sica, 
          Universidade Federal de Alagoas (UFAL), 
          Av.~Lourival Melo Mota, s/n, 57072-900 Macei\'o, AL, Brazil; and\\
    Instituto Tecnol\'ogico de Buenos Aires (ITBA),
          Av.~Eduardo Madero 399, C1106ACD Ciudad Aut\'onoma de Buenos Aires, Argentina; and\\
          Complex Systems Group, 
          Facultad de Ingenier\'{\i}a y Ciencias Aplicadas,
          Universidad de los Andes, 
          Av. Mons. \'Alvaro del Portillo 12.455, Las Condes, Santiago, Chile
\\
\bf{2} Centro de Ci\^encias Exatas e da Natureza, 
          Departamento de Estat\'istica,
          Universidade Federal de Pernambuco (UFPE), 
          Cidade Universit\'aria, 50740-540 Recife, PE, Brasil
\\
\bf{3} Instituto de Computa\c c\~ao, 
          Universidade Federal de Alagoas (UFAL), 
          Av.~Lourival Melo Mota, s/n, 57072-900, Macei\'o, AL, Brazil
\\
\bigskip

* oarosso@gmail.com

\end{flushleft}

\section*{Abstract}
We present a new approach for online handwritten signature classification and verification based on 
descriptors stemming from Information Theory.
The proposal uses the Shannon Entropy, the Statistical Complexity, and the Fisher Information evaluated over the Bandt and Pompe symbolization of the horizontal and vertical coordinates of signatures.
These six features are easy and fast to compute, and they are the input to an One-Class Support Vector Machine classifier.
The results produced surpass state-of-the-art techniques that employ higher-dimensional feature spaces which often require specialized software and hardware.
We assess the consistency of our proposal with respect to the size of the training sample, and we also use it to classify the signatures into meaningful groups.

\section*{Introduction}

The word {\it biometrics\/} is associated to human traits or behaviors which can be  measured 
and used for individual recognition.  
In fact, the biometry  recognition, as a personal authentication  signal processing, can 
be used in applications where users need to be security identified\cite{OrtegaGarcia2004}.  
Clearly,  these kind of systems can either verify or identify.  

Two types of biometrics  can be defined according to the personal traits considered:  
physical/physiological or behavioral. 
Physical/physiological biometrics is about catering the biological traits of users, like fingerprints, 
iris, face, hand, etc. 
Behavioral biometrics takes into account dynamic traits of users, such as, voice, handwritten and 
signature expressions. 

One of the main advantages of biometric systems is that users do not have 
to remember passwords or carry access keys. 
Another important advantage lies in the difficulty to steal, imitate or generate genuine 
biometric data, leading to enhanced security\cite{OrtegaGarcia2004}.

As mentioned,  behavioral biometrics is based on measurements extracted from an activity 
performed by the user, in conscious or unconscious  way, that are inherent to his/her 
own personality or learned behavior.
In this aspect,  behavioral biometrics has interesting pros, like user acceptance and 
cancelability, but it still lacks of some level of the uniqueness physiological biometrics has.  

Among the pure behavioral biometric traits, the handwritten 
signature and the way we sign is the one with widest social and legal acceptance\cite{Plamondon1989,Leclerc1994,Gupta1997,Impedovo2008,Ahmed2013}. 
Identity verification by signature analysis requires no invasive measurements and people 
are familiar with the use of signatures in their daily life.  
Also, it is the modality confronted with the highest level of attacks. 

A signature is a handwritten depiction of someone's name or some other mark of 
identification written on documents and devices as proof of identification. 
The formation of signature varies from person to person, or even from the same person due 
to the psychophysical state of the signer and the conditions under which the signature 
apposition process occurs.

Hilton\cite{Hilton1992} studied how signatures are produced, and found that the signature has at least three attributes: 
form, movement  and variation;  being  movement the most important, because signatures 
are produced by moving a writing device. 
The study also noted that a person's signature does evolve over time and, with the vast majority 
of users, once the signature style has been established the modifications are usually slight. 
The movement is produced by muscles of fingers, hand, wrist, and in some writers the arm; 
these muscles are controlled by nerve impulses.  
When one person is signing these nerve impulses are controlled by the brain without any 
particular attention to detail. 
The signing processes  can be described then, at high level, as how the central nervous 
system (the brain) recovers information from long term memory in which parameters such as 
size, shape, timing etc.  are specified.  
At the peripheral level, commands are generated for muscles. 
In consequence, the signing process is believed to be a reflex action 
(ballistic action\footnote{Ballistic movement can be defined as muscle contractions that exhibit maximum velocities and accelerations over a very short period of time. 
They exhibit high firing rates, high force production, and very brief contraction times\cite{ballistic}.}) 
rather than  a deliberate action. 
Then, the production of genuine signatures is associated to a ballistic handwriting, which 
is characterized by  a spurt of activity, without positional feedback, whereas  the 
production of forgery signature is associated to a deliberate handwriting which is 
characterized by a conscious attempt to produce a visual pattern with the aid of positional 
feedback\cite{DerGon1965,Nalwa1997}.

Handwritten signature verification is a problem in which the input signature 
(a test signature) is classified as genuine or forged. 
This process is usually performed in three main phases:\cite{Plamondon1989,Leclerc1994,Gupta1997,Impedovo2008,Ahmed2013} 
\begin{itemize}
   \item [$\bullet$]{\bf Data acquisition and pre-processing.}
   Two different categories of systems can be identified, depending on whether there is electronic 
   access to the handwritten process or not. 
   {\it a)~Online or dynamic recognition\/}, in which the pen's instantaneous information trajectories, 
   and also information like pressure, speed  or pen-up movements can be captured.
   {\it b)~Offline or static recognition\/}: those that record signatures as images 
   on paper which can be later digitized by means of a scanner, and processed.  
   In the latter, the pre-processing phase involves filtering, noise reduction and 
   smoothing.
   Online signature verification offers reliable identity protection, as it employs dynamic information not available on the signature image itself  but in the process of signing. As a consequence, 
   online signature verification systems usually achieve better accuracy than offline systems.

   \item [$\bullet$]{\bf Feature extraction.\/}  
   Two types of features can be used. 
   {\it a)~Function features of the signature\/}: time 
   functions whose values constitute the feature set. 
   {\it b)~Parameter features\/}: the signature is characterized as a vector of 
   elements, each one representative of the value of the feature. 
   Usually, the last one yields better performance, but it is also time-consuming.

   \item [$\bullet$]{\bf Classification.\/}
   In the verification process, the authenticity of the test signature is evaluated by matching 
   it against those stored in the knowledge base developed during the enrollment stage. 
   This process produces a single response that attests to the authenticity of the test signature. 
   When template matching techniques are considered, a questioned sample is matched against 
   templates of authentic/forgery signatures. 
   Distance-based classifiers, mostly when parameters are used as features, are usually developed 
   with statistical techniques, e.g. with Mahalanobis and Euclidean distances.
  The performance of a signature verification system is commonly assessed in terms 
   of the percentage Equal Error Rate.
\end{itemize}

On the one hand, template matching attempts at finding similarities between the input signature and those in a data base.
Most approaches use Dynamic Time Warping to perform this match\cite{Impedovo2008,Ahmed2013}.
On the other hand, distance-based classifiers rely on the use of features derived from the signatures.

Two opposite mechanisms describing the signing process can be found in the literature.
The nonlinear character and chaotic behavior of several physiological complex processes are well 
established\cite{Goldberger1990,West2013}. 
In particular, Longstaff and Heath\cite{Longstaff1999}
found evidence of chaotic behavior on the underlying dynamics
of time series related to velocity profiles of handwritten texts. 
Taking into account the inherent behavioral nature of the online signing process, the input 
information could be associated to deterministic (nonlinear low dimensional chaotic) signals, and 
the handwritten signature variations as a consequence of chaos (sensibility to initial conditions). 
In opposition, most of the research in the field of  signal verification considers  the input 
information as well described by a random  process\cite{Plamondon1989,Leclerc1994,Gupta1997,Impedovo2008,Ahmed2013}. 
Then, the dynamic input information acquired through a time sampling procedure must be 
consequently considered as discrete time random sequence. 
In any case, the signature analysis taken as a time-based sequence characterization process is 
strongly related to the way in which a reference model is established. 
From the stochastic point of view, Hidden Markov Models are among the most commonly used in the 
literature, and the ones with the best performance in signature 
verification\cite{Plamondon1989,Leclerc1994,Gupta1997,Impedovo2008,Ahmed2013}.

Our proposal relies on the use of time causal quantifiers based on Information Theory for the 
characterization  of online handwritten signatures: 
normalized permutation Shannon entropy, permutation statistical complexity and permutation Fisher 
information measure. 
These quantifiers have proved to be useful in the identification of chaotic and stochastic 
dynamics throughout the associated time series\cite{Rosso2007,Rosso2015}. 
Their evaluation is simple and fast, making them apt for the signature verification problem. 
We apply our proposal to the well know MCYT online signature data base\cite{MCYT2003}.

Next section describes the database used in this study, followed by a section where we detail 
the quantifiers employed and by their application to the data.
In addition to the usual data flow, we present an exploratory data analysis (EDA) of the features that enhances their appropriateness for this problem.
The expressiveness and usefulness of these descriptors for the problem of online signature 
classification and verification follows in the sequence: we experiment their application to the test-bed.

\section*{Handwritten signatures database}
\label{Sec:HandwrittenData}

The present study is carried out on the freely available and widely used handwritten signatures 
database MCYT-100 subset of 100 persons\cite{MCYT2003}.
The acquisition of each on-line signature is accomplished dynamically using a graphics tablet. 
The signatures  are acquired on a
WACOM$^\copyright$ graphic tablet, model~INTUOS~A6~USB.
The tablet resolution is  \SIunits{2540}~{lines\per in} (\SIunits{100}~{lines\per\milli\meter}), 
and the precision is \SIunits{$\pm$0.25}~{\milli\meter}. 
The maximum detection height is \SIunits{10}~{\milli\meter} 
(so also pen-up movements are considered), and the capture area is 
\SIunits{127}~{\milli\meter} (width) $\times$ \SIunits{97}~{\milli\meter} (height).
This tablet provides the following discrete-time sequences:
{\it a)\/} position $x_t$ in the $x$-axis,
{\it b)\/} position $y_t$ in the $y$-axis, and 
{\it c)\/} 
also the time series corresponding to the pressure $p_t$ applied by the pen, as well as the
azimuth  $\gamma_t$ and altitude $\varphi_t$ angles of the pen with respect to the tablet, not used in
the present work.
The sampling frequency is set to \SIunits{100}~{\hertz}.
Taking into account the Nyquist sampling criterion and the fact that the maximum frequencies of the
related  biomechanical sequences are always  under  \SIunits{20-30}~{\hertz}\cite{Baron1989}, 
this sampling frequency leads to a precise discrete-time signature representation.

\begin{figure}[hbt]
\centering
\includegraphics[width = \linewidth]{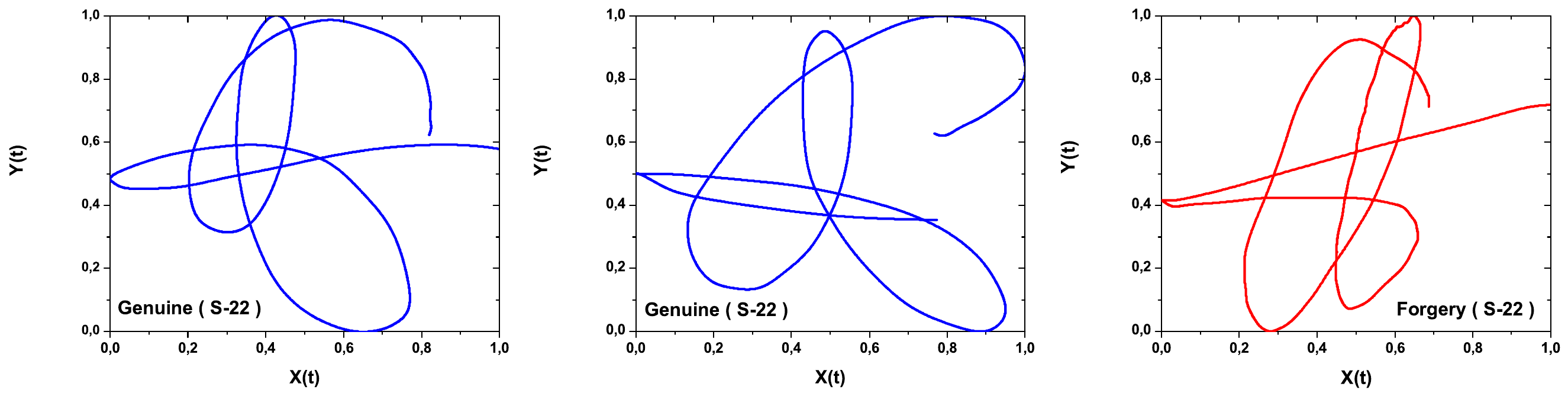}
\includegraphics[width = \linewidth]{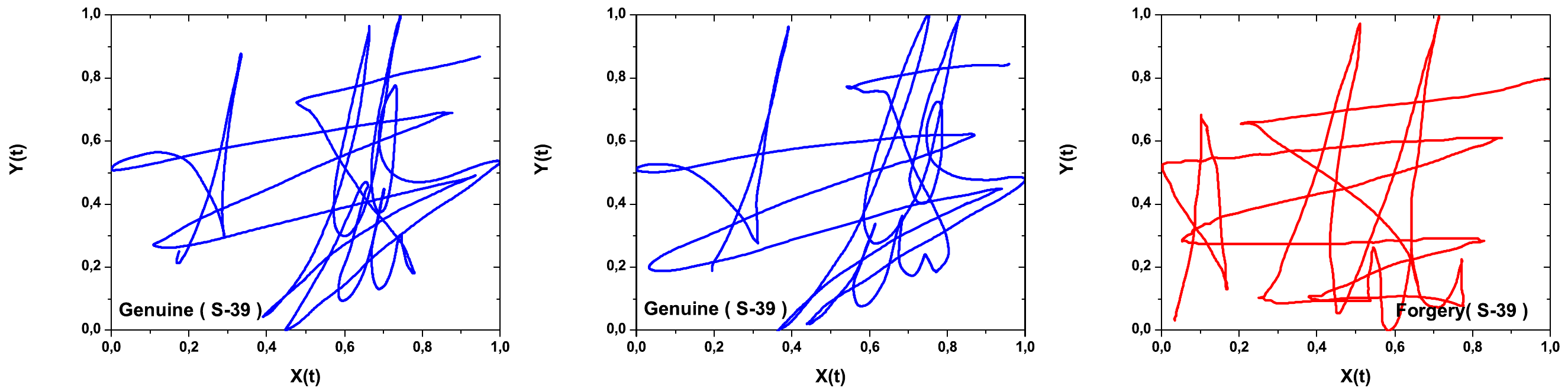}
\includegraphics[width = \linewidth]{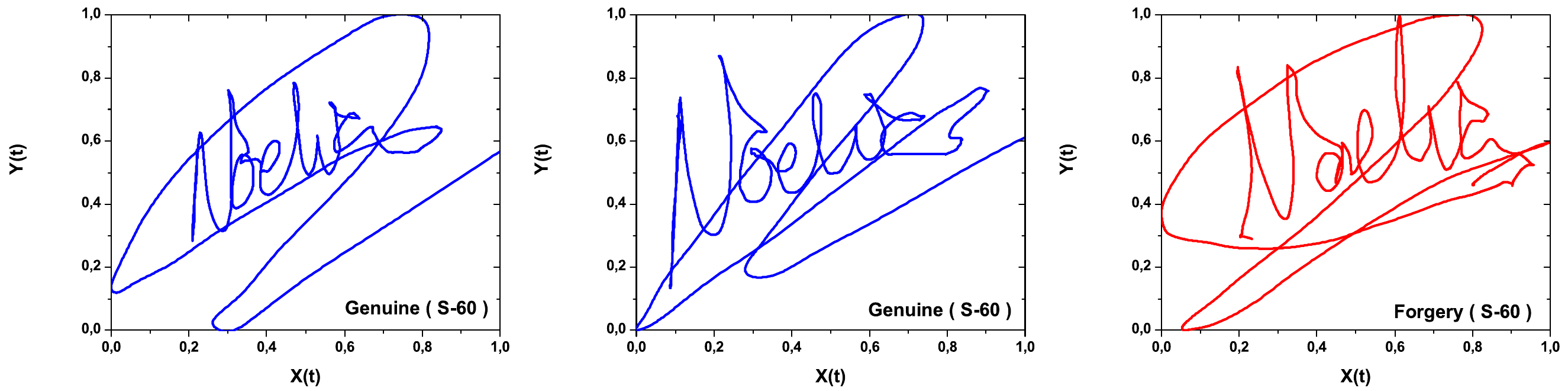}
\includegraphics[width = \linewidth]{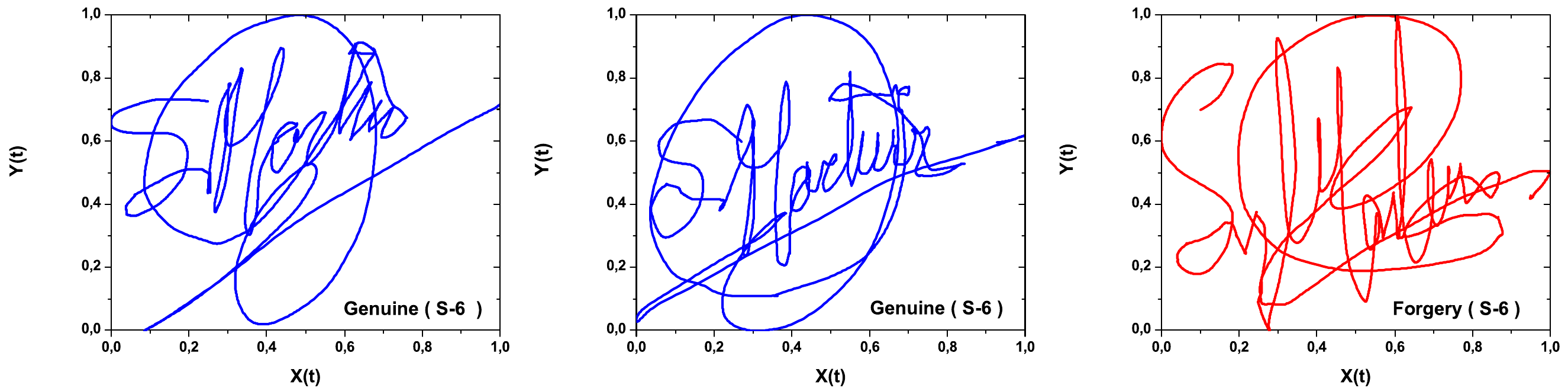}
\includegraphics[width = \linewidth]{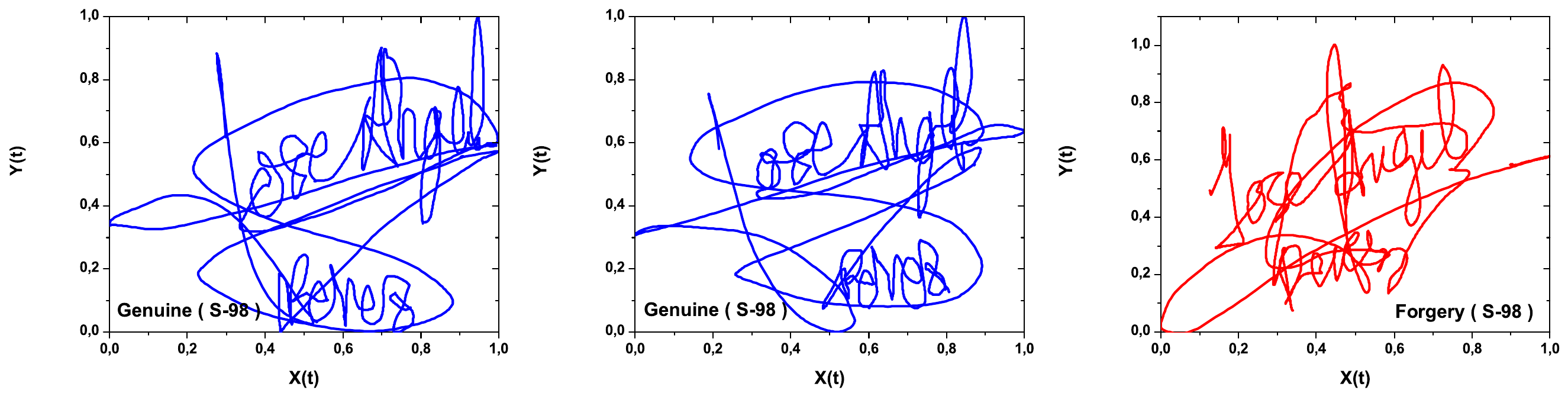}
\includegraphics[width = \linewidth]{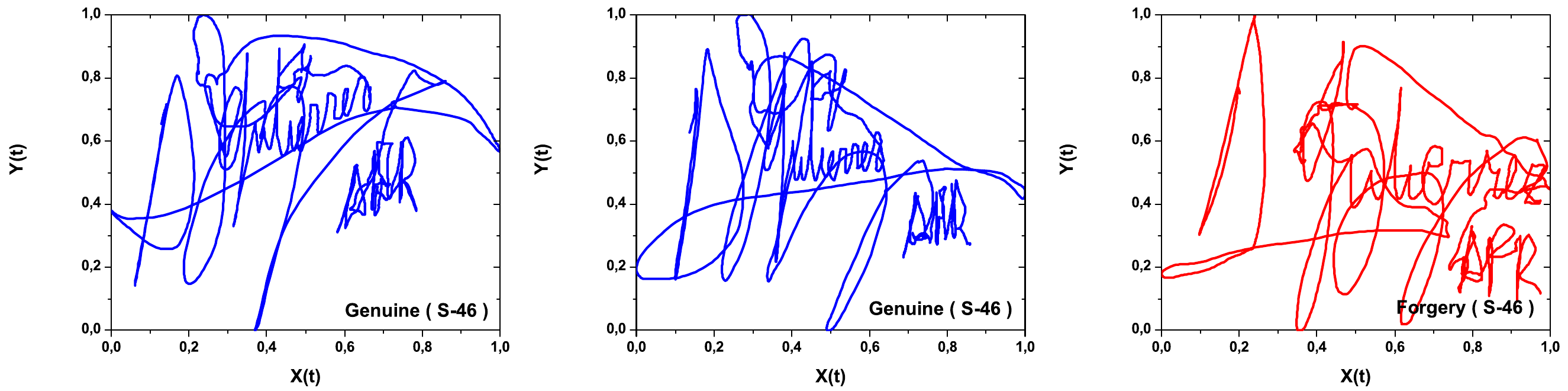}
\caption{Six different subjects signatures from the MCYT database.
Two genuine signatures (left, blue) and a skilled forgery (right, red). 
The two first signatures were classified as H1A and H1B,
the following two to types H2A and H2B, and the last two to types H3A and H3B; cf.\ Sec.\ Signatures classification.}
\label{fig:MCYT-firmas}
\end{figure}

The signature corpus comprises genuine and shape-based highly skilled forgeries with natural 
dynamics\cite{MCYT2003,Salicetti2009}.
In order to obtain the forgeries, each contributor is requested to imitate other signers by writing
naturally.
For this task, they were given the printed signature to imitate and were asked not only to imitate the
shape but also to generate the imitation without artifacts such as breaks or slow-downs (see 
\cite{MCYT2003,Salicetti2009} for more details of the acquisition procedure).
Each signer contributes with $25$ genuine signatures in five groups of five signatures each, 
and is forged $25$ times by five different imitators.
Figure~\ref{fig:MCYT-firmas} presents examples for six different subjects,
being the first two columns genuine and the third column forgery signatures.

Since signers are concentrated in a different writing task between genuine signature sets, the 
variability between client signatures from different acquisition sets is expected to be  higher than 
the variability of signatures within the same set.
The total number of contributors in the MCYT is $330$, and the total number of signatures present in
the signature database is $16,500$, half of them genuine signatures and the rest 
forgeries\cite{MCYT2003,Salicetti2009}.

As previously mentioned, we used a subset of the database, denominated MCYT-100, 
which includes $100$ subjects and for each one, $25$ genuine  and $25$ 
skilled  
forged signatures, and only the 
corresponding time series corresponding to the $x$- and $y$-coordinates of each signature will be analyzed.
In particular, one must note that the time series' lengths are quite variable.
In order to facilitate our Information Theory analysis, we pre-processed each time series as follows:
{\it a)\/}~the coordinates were re-scaled into the unit square $[0,1]\times[0,1]$; 
{\it b)\/}~taken as base these scaled values, the original total number of data for 
each time series is expanded to $M = 2000$ points using a cubic Hermite polynomial.
In this way, for each subject $k$ ($k = 1, \dots, 100$) and associated signatures $j$ ($j=1, \dots, 25$) 
we will analyze two time series, denoted by
$ {\mathbf X}^{(k;\alpha)}_j = \{ 0 \leq {\tilde{x}}^{(k;\alpha)}_{j;i} \leq 1,~i = 1, \ldots, M \}$  and 
$ {\mathbf Y}^{(k;\alpha)}_j = \{ 0 \leq {\tilde{y}}^{(k;\alpha)}_{j;i} \leq 1,~i = 1, \ldots, M \}$,
in which the supra-index $\alpha = G,~F$ denotes genuine and forgery signature, and 
${\tilde{x}}$ and ${\tilde{y}}$  are the interpolated values, respectively.

\section*{Information Theory quantifiers}
\label{Sec:Quantifiers}

The basic elements for the study of a system dynamics, either natural or man-made, are sequences of measurements or observations whose evolution can be tracked through time.
Then, given
an observable of such system, a natural question that arises is: how much information is this 
observable encoding about the dynamics of the underlying system?
The information contents of a system are typically evaluated via a probability distribution function (PDF) $P$ 
obtained from such observable. 
We can define Information Theory quantifiers as measures able to characterize relevant properties of the 
PDF
associated with these time series, and in this way we should judiciously 
extract information on the dynamical system under study. 

\subsection*{Shannon entropy, Fisher Information Measure, and Statistical Complexity}
\label{Sec:HFC}

Entropy is a basic quantity with multiple field-specific interpretations; for instance, it has been associated 
with disorder, state-space volume, and lack of information\cite{Brissaud2005}.
When dealing with information content, the Shannon entropy is often considered the foundational and most 
natural one\cite{Shannon1948,Shannon1949}. 

Given a continuous probability distribution function (PDF) $f(x)$ with $x \in \Omega \subset {\mathbb R}$ and 
$\int_{\Omega} f(x)~dx = 1$, its associated {\it Shannon Entropy\/} $S$  \cite{Shannon1948,Shannon1949} is
\begin{equation}
\label{shannon}
{\mathrm S}[f] = -\int_{\Omega} f(x) \ln f(x)  dx .
\end{equation}
It is a global measure, that is, it is not too sensitive to strong changes in the distribution taking 
place on a small-sized region of $\Omega$. 
Such is not the case with {\it Fisher's Information Measure\/} (FIM) $\mathcal F$\cite{Fisher1922,Frieden2004}, 
which constitutes a measure of the gradient content of the distribution $f$, thus being quite sensitive even 
to tiny localized perturbations. 
It reads
\begin{equation}
\label{fisher}
{\mathcal F}[f] = \int_{\Omega} { \frac{1}{f(x)} } \left[ { \frac{df(x)}{dx} }\right]^2 dx
    = 4 \int_{\Omega} \left[ { \frac{d \psi(x)} {dx} }\right]^2 ,\quad
    \text{where } \psi(x) = \sqrt{f(x)}.
\end{equation}

The Fisher Information Measure can be variously interpreted as a measure of the ability to estimate a parameter,
as the amount of information that can be extracted from a set of measurements, and also as a measure of the state
of disorder of a system or phenomenon\cite{Frieden2004}, its most important property being the so-called
Cramer-Rao bound.
It is important to remark  that the gradient operator significantly influences the contribution of minute local
$f$-variations to the Fisher information value, accordingly, this quantifier is called ``local"\cite{Frieden2004}. 
Note that the Shannon entropy decreases with the distribution skewness, while the Fisher information increases.

Local sensitivity is useful in scenarios whose description necessitates an appeal to a notion of ``order''.
In the previous definition of FIM (Eq.~(\ref{fisher})) the division by $f(x)$ is not convenient  if 
$f(x) \rightarrow 0$ at certain points of the support $\Omega$. 
We avoid this if we work with real probability amplitudes, by means of the alternative expression that employs 
$\psi(x)$\cite{Fisher1922,Frieden2004}.
This form requires no divisions, and shows that $\mathcal F$ simply measures the gradient content in 
$\psi(x)$.

Let now $P=\{p_i; i=1,\ldots, N\}$  be a  discrete probability distribution, with $N$ the number of possible 
states of the system under study.
The Shannon's logarithmic information measure reads
\begin{equation}
\label{shannon-disc}
{\mathrm S}[P] = -\sum_{i=1}^{N} p_i \ln p_i.  
\end{equation}
This can be regarded to as a measure of the uncertainty associated (information) to the physical process described by $P$.
For instance, if ${\mathrm S}[P] = {\mathrm S}_{\min} = 0$, we are in position to predict with complete certainty 
which of the possible outcomes $i$, whose probabilities are given by $p_i$, will actually take place. 
Our knowledge of the underlying process described by the probability distribution is maximal in this instance. 
In contrast, our knowledge is minimal for a uniform distribution $P_e = \{ p_i = 1/N, \forall i=1, \ldots , N \}$
since every outcome exhibits the same probability of occurrence, and the uncertainty is maximal, i.e.,
 ${\mathrm S}[P_e] = {\mathrm S}_{\max} = \ln N$.
In the discrete case,  we define a ``normalized" Shannon entropy, $0 \leq {\mathcal H} \leq 1$, as
\begin{equation}
\label{shannon-disc-normalizada}
{\mathcal H} [P] = {\mathrm S}[P]  / {\mathrm S}_{\max} \ .
\end{equation}

The concomitant problem of loss of information due to the discretization has been thoroughly studied (see, for
instance, \cite{Zografos1986,Pardo1994} and references therein) and, in particular, it entails the
loss of Fisher's shift-invariance, which is of no importance for our present purposes.
For the FIM we take the expression in terms of  real probability amplitudes as starting point, then a discrete 
normalized FIM,  $0 \leq {\mathcal F} \leq 1$, convenient for our present purposes, is given by
\begin{equation}
\label{Fisher-disc}
{\mathcal F}[P]=F_0\sum_{i=1}^{N-1} \big[\sqrt{p_{i+1}} - \sqrt{p_{i}}\big]^2 .
\end{equation}
It has been extensively discussed that this discretization is the best behaved in a  discrete environment\cite{Dehesa2009}. 
Here the normalization constant $F_0$ reads
\begin{equation}
\label{F0}
F_0=\left\{
       \begin{array}{rl}
                   1,       & \text{if } p_{i^*} = 1 \text{ for }
                            i^* = 1 \text{ or } i^* = N \text{ and } p_{i}  = 0, \forall  i \neq i^*, \\
                    1/2,     & \text{otherwise.}
       \end{array}
\right.
\end{equation}

The perfect crystal and the isolated ideal gas are two typical examples of systems with minimum and 
maximum entropy, respectively. 
However, they are also examples of simple models and therefore of systems with zero complexity, as the
structure of the perfect crystal is completely described by minimal information (i.e., distances and 
symmetries that define the elementary cell) and the probability distribution for the accessible states 
is centered around a prevailing state of perfect symmetry. 
On the other hand, all the accessible states of the ideal gas occur with the same probability and can be
described by a ``simple" uniform distribution. 

According to L\'opez-Ruiz {\it et al.}\cite{LMC1995}, and using an oxymoron, an object, a procedure, 
or system is said to be complex when it does not exhibit patterns regarded as simple. 
It follows that a suitable complexity measure should vanish both for completely ordered and for completely 
random systems and cannot only rely on the concept of information (which is maximal and minimal for the 
above mentioned systems).
A suitable measure of complexity can be defined as the product of a measure of information and a measure of
disequilibrium, i.e. some kind of distance from the equiprobable distribution of the accessible states of 
a system. 
In this respect, Rosso and coworkers\cite{Lamberti2004} introduced an effective {\it Statistical Complexity 
Measure\/} (SCM) ${\mathcal C}$, that is able to detect essential details of the dynamical processes 
underlying the dataset.

Based on the seminal notion advanced by L\'opez-Ruiz {\it et al.}\cite{LMC1995}, this statistical complexity 
measure\cite{Lamberti2004} is defined through the product
\begin{equation}
{\mathcal C}[P] = {\mathcal Q}_{J}[P,P_e] \cdot {\mathcal H}[P]
\label{complexity}
\end{equation}
of the normalized Shannon entropy ${\mathcal H}$, see Eq.~\eqref{shannon-disc-normalizada}, and the disequilibrium 
${\mathcal Q}_{J}$ defined in terms of the Jensen-Shannon divergence ${\mathcal J}[ P, P_e]$.
That is,
\begin{equation}
\label{disequilibrium}
{\mathcal Q}_{J} [ P, P_e] = Q_{0} {\mathcal J}[ P, P_e] = 
Q_{0} \{ {\mathrm S}[(P + P_e)/2 ] - {\mathrm S}[ P ]/2 - {\mathrm S}[P_e]/2\},
\end{equation}
the above-mentioned Jensen-Shannon divergence and $Q_0$, a normalization constant 
such that $0 \leq {\mathcal Q}_{J} \leq 1$:
\begin{equation}
Q_0 = -2 \left\{  {\frac{N+1}{N}}  \ln (N+1) - \ln (2N)  +  \ln N \right\}^{-1} \ ,
\label{q0-jensen-1}
\end{equation}
are equal to the inverse of the maximum possible value of ${\mathcal J} [P,P_e]$.
This value is obtained when one of the components of $P$, say $p_m$, is equal to one and the remaining $p_j$ 
are zero.

The Jensen-Shannon divergence, which quantifies the difference between probability distributions, 
is especially useful to compare the symbolic composition between different sequences\cite{Grosse2002}.
Note that the above introduced 
SCM
depends on two different probability distributions: one associated with the system under analysis, $P$, and the other the uniform distribution, $P_e$.
Furthermore, it was shown that for a given value of ${\mathcal H}$, the range of possible ${\mathcal C}$ 
values varies between a minimum ${\mathcal C}_{min}$ and a maximum ${\mathcal C}_{max}$, 
restricting the possible values of the 
SCM\cite{Martin2006}.

Thus, it is clear that important additional information related to the correlational structure between the 
components of the physical system is provided by evaluating the statistical complexity measure. 
In this way, the information plane ${\mathcal H} \times {\mathcal C}$ constitute a nice tool to visualizate 
and characterize different dynamical systems.

If our system lies in a very ordered state, which occurs when almost all the $p_{i}$--values are zeros 
except for a particular state $k \neq i$ with $p_{k} \cong 1$, both
the normalized Shannon entropy and statistical complexity are close to zero (${\mathcal H} \approx 0$ and 
${\mathcal C} \approx 0$), and the normalized Fisher's information measure is close to  one (${\mathcal F} \approx 1$).
On the other hand, when the system under study is represented by  a very disordered state, that is when all the 
$p_{i}$--values oscillate around the same value, we have ${\mathcal H} \approx 1$ while 
${\mathcal C} \approx 0$ and ${\mathcal F} \approx 0$.
One can state that the general FIM--behavior of the present discrete version 
(Eq.~(\ref{Fisher-disc})),  
is opposite to that of the Shannon entropy, except for periodic motions.
The local sensitivity of FIM for discrete--PDFs is reflected in the fact that the specific ``$i-$ordering" 
of the discrete values $p_{i}$ must be seriously taken into account in evaluating the sum in
 Eq.~(\ref{Fisher-disc}).
This point was extensively discussed by Rosso 
and co-workers\cite{Olivares2012A,Olivares2012B}.
The summands can be regarded to as a kind of ``distance" between  two contiguous probabilities.
Thus, a different ordering of the pertinent summands would lead to a different FIM-value, hereby its local nature.
In the present work, we follow the Lehmer lexicographic order\cite{Lehmer} in the generation 
of Bandt and Pompe PDF (see next section).
Given the local character of FIM, when combined with a global quantifier as the normalized Shannon entropy, 
conforms the Shannon--Fisher plane, ${\mathcal H} \times {\mathcal F}$, introduced by Vignat and Bercher\cite{Vignat2003}. 
These authors showed that this plane is able to characterize the non-stationary behavior of a complex signal.
 
\subsection*{The Bandt and Pompe approach to the PDF determination}
\label{Sec:Bandt-Pompe}

The evaluation of the Information Theory derived quantifiers, like those previously introduced (Shannon entropy,
Fisher information and statistical complexity), suppose some prior knowledge about the system; specifically, 
a probability distribution associated to the time series under analysis should be provided beforehand. 
The determination of the most adequate PDF is a fundamental problem because the PDF $P$ and the sample space 
$\Omega$ are inextricably linked. 

Usual methodologies assign to each time point of the series ${\mathcal X}$ a symbol from a  finite alphabet 
$\mathfrak{A}$, thus creating a {\it symbolic sequence} that can be regarded to as a {\it non causal coarse grained} 
description of the time series under consideration. 
As a consequence, order relations and the time scales of the dynamics are lost. 
The usual histogram technique corresponds to this kind of assignment.
{\it Causal information\/}  may be duly incorporated if information about the past dynamics of the system is 
included in the symbolic sequence, i.e., symbols of alphabet $\mathfrak{A}$ are assigned to a portion of the 
phase-space or trajectory.

Many methods have been proposed for a proper selection of the probability space $(\Omega, P)$. 
Bandt and Pompe (BP)\cite{Bandt2002} introduced a simple and robust symbolic methodology that takes into account 
time causality of the time series (causal coarse grained methodology) by comparing neighboring values in a 
time series.
The symbolic data are:
{\it (i)\/}~created by ranking the values of the series; and
{\it (ii)\/}~defined by reordering the embedded data in ascending order, which is tantamount to a phase space 
reconstruction with embedding dimension (pattern length) $D$ and time lag $\tau$.
In this way, it is possible to quantify the diversity of the ordering symbols (patterns) derived from a scalar 
time series.

Note that the appropriate symbol sequence arises naturally from the time series, and no model-based assumptions 
are needed.
In fact, the necessary ``partitions'' are devised by comparing the order of neighboring relative values rather 
than by apportioning amplitudes according to different levels.
This technique, as opposed to most of those in current practice, takes into account the temporal structure of the time series generated by the physical process under study.
As such, it allows us to uncover important details concerning the ordinal structure of the time 
series\cite{Rosso2007,Rosso2012} and can also yield information about temporal correlation\cite{Rosso2009A,Rosso2009B}.

It is clear that this type of analysis of a time series entails losing details of the original series' 
amplitude information.
Nevertheless, by just referring to the series' intrinsic structure, a meaningful difficulty reduction has 
indeed been achieved by BP with regard to the description of complex systems.
The symbolic representation of time series by recourse to a comparison of consecutive ($\tau = 1$) or 
nonconsecutive ($\tau > 1$) values allows for an accurate empirical reconstruction of the underlying phase-space, 
even in the presence of weak (observational and dynamic) noise\cite{Bandt2002}.
Furthermore, the ordinal patterns associated with the PDF are invariant with respect to nonlinear monotonous 
transformations.
Accordingly, nonlinear drifts or scaling artificially introduced by a measurement device will not modify the 
estimation of quantifiers, a nice property if one deals with experimental data (see, e.g.,\cite{Saco2010}).
These advantages make the BP methodology more convenient than conventional methods based on range 
partitioning, i.e., a PDF based on histograms.

To use the BP methodology\cite{Bandt2002} for evaluating the PDF, $P$, associated with the time 
series (dynamical system) under study, one starts by considering partitions of the $D$-dimensional space that 
will hopefully ``reveal'' relevant details of the ordinal structure of a given one-dimensional time series 
${\mathcal X}(t) = \{ x_t; t = 1, \ldots, M\}$ with embedding dimension $D > 1$ ($D \in {\mathbb N}$) 
and time lag $\tau$ ($\tau \in {\mathbb N}$).
We are interested in ``ordinal patterns'' of order (length) $D$ generated by
\begin{equation}
\label{asignation1}
(s)\mapsto \left(x_{s-(D-1)\tau},x_{s-(D-2)\tau},\ldots, x_{s-\tau},x_{s}\right)  ,
\end{equation}
which assign to each time $s$ the $D$-dimensional vector of values at times $s, s-\tau,\ldots,s-(D-1)\tau$.
Clearly, the greater $D$, the more information on the past is incorporated into our vectors.
By ``ordinal pattern'' related to the time $(s)$, we mean the permutation $\pi=(r_0,r_1, \ldots,r_{D-1})$ 
of $[0,1,\ldots,D-1]$ defined by 
\begin{equation}
\label{asignation2}
x_{s-r_{D-1}\tau} \le~x_{s-r_{D-2}\tau} \le \cdots \le~x_{s-r_{1}\tau} \le x_{s-r_0\tau}  .
\end{equation}
We set $r_i < r_{i-1}$ if $x_{s-r_{i}} = x_{s-r_{i-1}}$ for uniqueness, although ties in samples from continuous 
distributions have null probability.

For all the $D!$ possible orderings (permutations) $\pi_i$ when  embedding dimension is $D$, 
and time-lag $\tau$,
their relative 
frequencies can be naturally computed according to the number of times this particular order sequence is found 
in the time series,  divided by the total number of sequences,
\begin{equation}
\label{eq:frequ}
p(\pi_i)= \frac{\# \{s|s\leq N-(D-1)\tau ; (s)  \text{ is of type } \pi_i \}}{N-(D-1)\tau} ,
\end{equation}
where $\#$ denotes cardinality.
Thus, an ordinal pattern probability distribution $P = \{ p(\pi_i), i = 1, \dots, D! \}$ is obtained from the 
time series.

\begin{figure}[hbt]
\centering
\includegraphics[width = \linewidth]{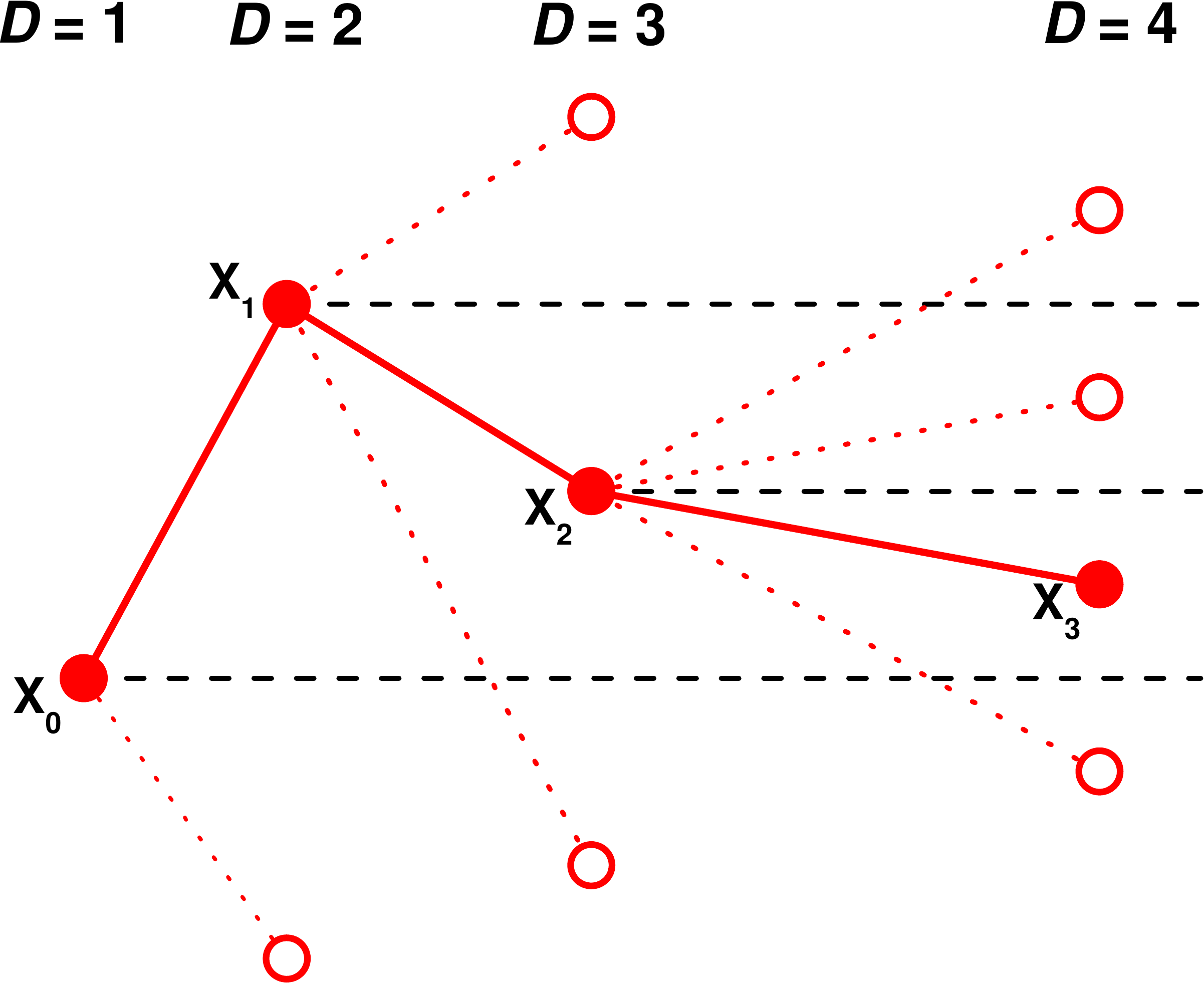}
\caption{Illustration of the construction principle for ordinal patterns of length $D$ \cite{Parlitz2012}. 
If $D=4$ and $\tau=1$, full circles and continuous lines represent the sequence of values $x_0 < x_1 > x_2 > x_3$ 
which lead to the pattern $\pi=[0321]$.}
\label{fig:patrones}
\end{figure}

Figure~\ref{fig:patrones} illustrates the construction principle of the ordinal patterns of length 
$D=2$, $3$ and $4$ with $\tau = 1$\cite{Parlitz2012}.
Consider the sequence of observations $\{x_0, x_1, x_2, x_3\}$.
For $D=2$, there are only two possible directions from $x_0$ to $x_1$: up and down.
For $D=3$, starting from $x_1$ (up) the third part of the pattern can be above $x_1$, below $x_0$, or between 
$x_0$ and $x_1$.
A similar situation can be found starting from $x_1$ (down). 
For $D=4$, for each one of the six possible positions for $x_2$, there are four possible localizations for $x_3$, 
yielding $D!=4!=24$ different possible ordinal patterns.
In Fig.~\ref{fig:patrones}, full circles and continuous lines represent the sequence values 
$x_0 < x_1 > x_2 > x_3$, which leads to the pattern  $\pi=[0321]$.
A graphical representation of all possible patterns corresponding to $D = 3, 4$ and $5$
can be found in Fig.~2 of Parlitz \textit{et al.}\cite{Parlitz2012}.

The embedding dimension $D$ plays an important role in the evaluation of the appropriate probability 
distribution, because $D$ determines the number of accessible states $D!$ and also conditions the minimum acceptable 
length $M \gg D!$ of the time series that one needs in order to work with reliable statistics\cite{Kowalski2007}.
Regarding the selection of the parameters, Bandt and Pompe suggested working with $4 \leq D \leq 6$, and specifically 
considered a time lag $\tau = 1$ in their cornerstone paper\cite{Bandt2002}.
Nevertheless, it is clear that other values of $\tau$ could provide additional information.
It has been recently shown that this parameter is strongly related, if it is relevant, to the intrinsic time scales 
of the system under analysis\cite{Zunino2010B,Soriano2011,Zunino2012}.

Additional advantages of the  method reside in
{\it i)\/} its simplicity (it requires few parameters: the pattern length/embedding dimension $D$ and the time lag 
$\tau$), and
{\it ii)\/} the extremely fast nature of the calculation process. 
The BP methodology can be applied not only  to time series representative of low dimensional dynamical systems, 
but also to any type of time series (regular, chaotic, noisy, or reality based).
In fact, the existence of an attractor in the $D$-dimensional phase space in not assumed.
The only condition for the applicability of the BP method is  a very weak stationary assumption: for $k \leq D$, 
the probability for $x_t < x_{t+k}$ should not depend on $t$.
For a review of BP's methodology and its applications to physics, biomedical and econophysics signals see 
Zanin {\it et al.\/}\cite{Zanin2012}. 
Moreover, 
Rosso {\it et al.\/}\cite{Rosso2007} show that the above mentioned quantifiers produce better descriptions 
of the process associated dynamics when the PDF is computed using BP rather than using the usual histogram methodology.

The BP proposal for associating probability distributions to time series (of an underlying symbolic 
nature) constitutes a significant advance in the study of nonlinear dynamical systems\cite{Bandt2002}.
The method provides univocal prescription for ordinary, global entropic quantifiers of the Shannon-kind.
However, as was shown by Rosso and coworkers\cite{Olivares2012A,Olivares2012B}, ambiguities arise in applying the 
BP technique with reference to the permutation of ordinal patterns. 
This happens if one wishes to employ the BP-probability density to construct local entropic quantifiers, 
like the Fisher information measure, which would characterize time series generated by nonlinear dynamical systems.

The local sensitivity of the Fisher information measure for discrete PDFs is reflected in the fact that the specific 
``$i$-ordering'' of the discrete values $p_i$ must be seriously taken into account in evaluating Eq.~(\ref{Fisher-disc}).
The numerator can be regarded to as a kind of ``distance'' between two contiguous probabilities.
Thus, a different ordering of the summands will lead, in most cases, to a different Fisher information value.
In fact, if we have a discrete PDF given by $P = \{ p_i, i = 1, \ldots , N\}$, we will have $N!$ possibilities 
{for the $i$-ordering.}

The question is, which is the arrangement that one could regard as the ``proper'' ordering?
The answer is straightforward in some cases, the histogram-based PDF constituting a conspicuous example.
For such a procedure, one first divides the interval $[a, b]$ (with $a$ and $b$ the minimum and maximum amplitude 
values in the time series) into a finite number on non-overlapping sub-intervals (bins).
Thus, the division procedure of the interval $[a, b]$ provides the natural order sequence for the evaluation of 
the PDF gradient involved in the Fisher information measure.
In our current paper, we chose the lexicographic ordering given by the algorithm of Lehmer\cite{Lehmer}, 
among other possibilities, due to its better distinction of different 
dynamics in the Shannon--Fisher plane, ${\mathcal H} \times {\mathcal F}$ (see\cite{Olivares2012A,Olivares2012B}).

\section*{Signature features and exploratory data analysis}
\label{Sec:Results}

Online handwritten classification and verification is an interesting and challenging classification problem.
On the one hand, intra-personal variation information can be large. 
Some people provide signatures with poor consistency. 
The speed, pressure and inclination, for example, pertaining to the signatures made by the same person can 
differ greatly  on regularity which makes it quite challenging to extract consistent features. 
On the other hand, we can only obtain few samples from one person and no forgeries in practice. 
This makes it very difficult to determine the reliability of extracted features. 
The main idea is to construct an efficient classification scheme for data acquisition, or the reduction of 
often unmanageable large datasets to a parsimonious form, without mislay important statistical information.
We aim at discovering relevant characteristic statistical structures which could be exploited if the key 
information can be efficiently condensed into a suitable low-dimensional object. 

The features we employ in this work are the Information Theory quantifiers already presented.
For each of the $k$ subjects ($k=1,\ldots,100$) in the database and its $j$ associated signatures 
($25$ genuine and $25$ 
skilled
forgery), two associated time series ${\mathbf X}^{(k;\alpha)}_j$ and 
${\mathbf Y}^{(k;\alpha)}_j$ are extracted and transformed into BP's PDFs with pattern length 
(embedding dimension) $D = 5$ and time lag $\tau = 1$.
Note that the condition $M \gg D!$ its satisfied.

We denoted these PDFs as:
\begin{align*}
P_{X;j}^{(k;\alpha)}&= \text{ Bandt and Pompe's PDF~of } {\mathbf X}^{(k;\alpha)}_j |_{D,\tau}, \text{ and}\\
P_{Y;j}^{(k;\alpha)}&= \text{ Bandt and Pompe's PDF of } {\mathbf Y}^{(k;\alpha)}_j |_{D,\tau},
\end{align*}
in which $j=1, \ldots, 25$, and $\alpha = G, F$ identify genuine and skilled forgery signatures, respectively.

We computed the normalized permutation Shannon entropy ${\mathcal H}$,
the permutation statistical complexity ${\mathcal C}$,
and the permutation Fisher information measure 
${\mathcal F}$ from these PDFs, and the obtained values are denoted as:
\begin{align*}
 {\mathcal H}_{X;j}^{(k;\alpha)} &=  {\mathcal H}[P_{X;j}^{(k;\alpha)}],  &{\mathcal H}_{Y;j}^{(k;\alpha)} &= {\mathcal H}[P_{Y;j}^{(k;\alpha)}];\\
 {\mathcal C}_{X;j}^{(k;\alpha)} &= \ {\mathcal C}[P_{X;j}^{(k;\alpha)}],  &{\mathcal C}_{Y;j}^{(k;\alpha)}  &= {\mathcal C}[P_{Y;j}^{(k;\alpha)}];\\
 {\mathcal F}_{X;j}^{(k;\alpha)} &= {\mathcal F}[P_{X;j}^{(k;\alpha)}],  &{\mathcal F}_{Y;j}^{(k;\alpha)} &= {\mathcal F}[P_{Y;j}^{(k;\alpha)}].
\end{align*}

We perform Exploratory Data Analysis (EDA)
on the Information Theory quantifiers 
looking for simple descriptions of the data. 
Apart from simple descriptive univariate measures, we use the Pearson correlation to measure the association between features. 
This analysis was performed using the {\tt R} language and 
platform version~3.2.1 (\url{http:\\www.R-project.org}).  

\begin{figure}[hbt]
\centering
\includegraphics[width=0.9\linewidth]{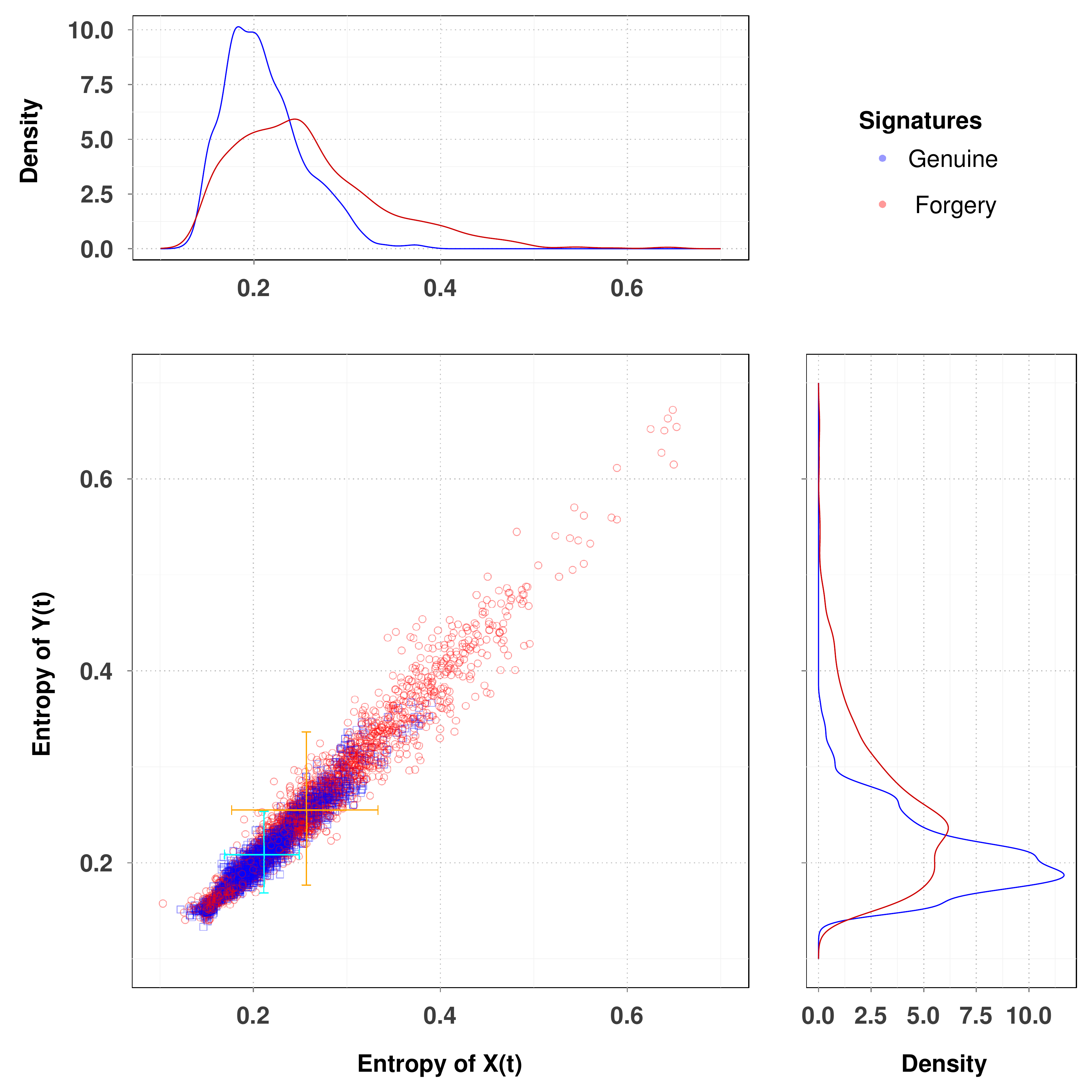}
\caption{Scatter plot with marginal  kernel density estimates of entropy quantifiers in both trajectory coordinates 
time series 
${\mathbf X}$ and ${\mathbf Y}$.
Genuine (blue) and skilled forgery signatures (red points), 100 subjects. 
Marginal kernel densities depict the distribution of entropy quantifiers along both axes.
}\label{Fig:HistEntropy}
\end{figure}

Figure~\ref{Fig:HistEntropy} shows a scatterplot of the Entropy for both the genuine and skilled forgery signatures.
The $5000$ points correspond to $25$ genuine signatures (in blue) and $25$ forgery signatures 
(in red) for each of the $100$ subjects. 
Both types of signatures show similar association (Correlation): 
${\rm Corr}({\mathcal H}_{X;j}^{(k;G)}, {\mathcal H}_{Y;j}^{(k;G)}) = 0.9665$ and 
${\rm Corr}({\mathcal H}_{X;j}^{(k;F)}, {\mathcal H}_{Y;j}^{(k;F)}) = 0.9770$.
The 
entropies of both types of signatures are overlapped and scattered elliptically. 
However, the bivariate mean and dispersion values differ. 

Entropies are less dispersed in the genuine than in the skilled forgery signatures,  
a signal of the separability between them. 
Marginal density plots show the distribution of entropy for each coordinate of both types of signatures. 
These plots, however limited due to its marginal nature, reveal several modes, and suggest both wide and narrow 
structures in the data. 

\begin{figure}[hbt]
\centering
\includegraphics[width=.95\linewidth]{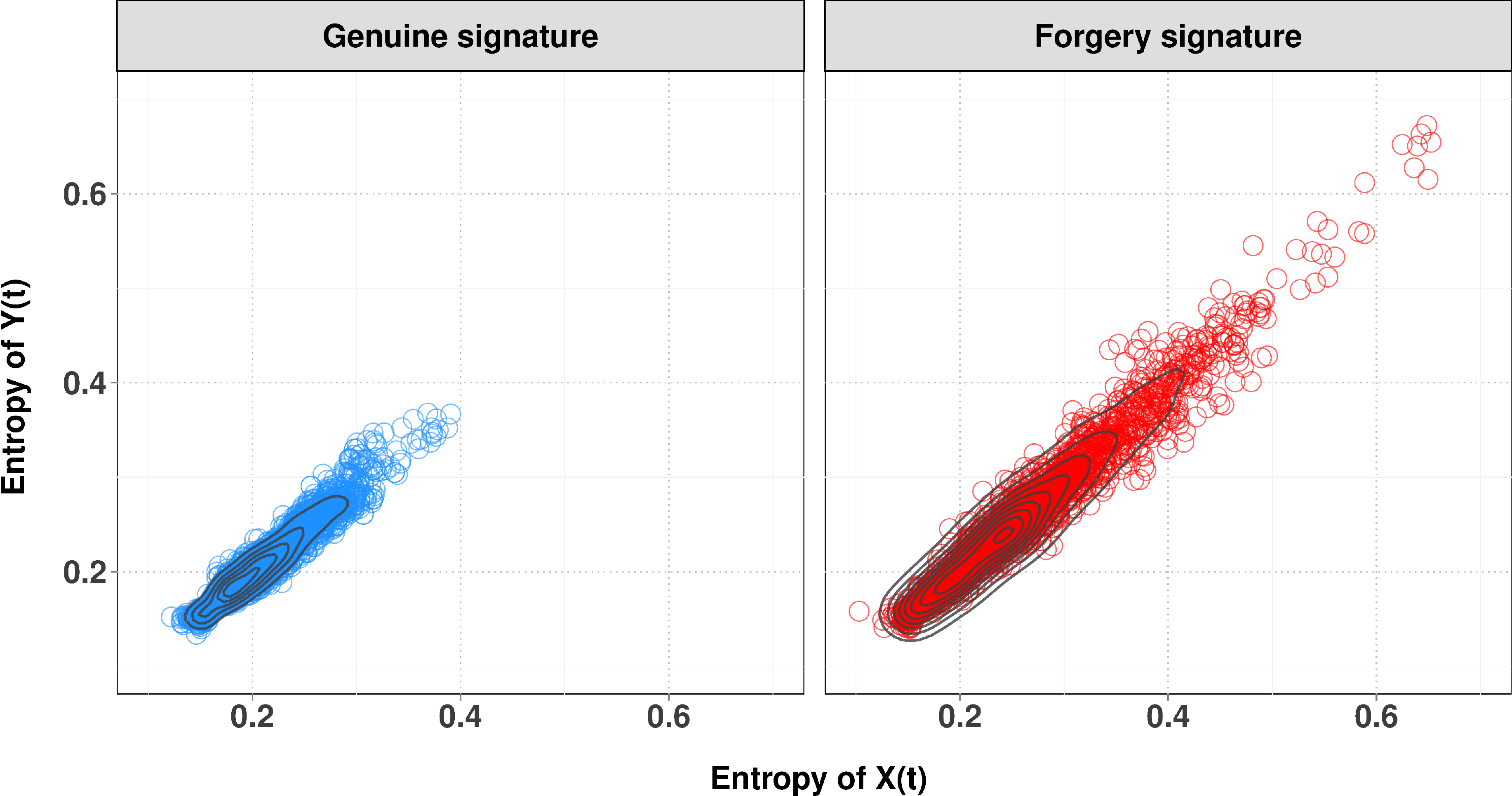}
\caption{Contour plot superimposed on the scatterplot of entropy quantifiers for genuine (right panel) 
and skilled forgery signatures (left panel)}
\label{Fig:ContourEntropy}
\end{figure}

Figure~\ref{Fig:ContourEntropy} shows the contour plots of bivariate kernel density estimates for the 
entropy in genuine and forgery signatures. 
A number of features are immediately noticeable. 
The dispersion in the former group is much smaller than in the latter (less than $0.4$). 
The kernel density estimates reveal skewness and a mild multimodality in the joint distribution of the data.
There are also quite many points that are far from these curves and cluster centers. 
These points correspond to abnormal local estimates obtained in heterogeneous blocks, possibly induced by the presence of clusters. 
The modes in genuine signatures are smaller than in forgery signatures, and this may be used as discriminatory measure. 
Similar results are obtained for the Complexity and the Fisher information; 
these are 
reported in the Supplementary Information, see Figs.~S1 to~S4, respectively.

\section*{Signatures classification}
\label{Sec-Classification}

As pointed out by Boul\'etreal {\it et al.\/}\cite{Bouletreau1998}, a signature  is characterized
by two aspects: 
{\it a)\/}~a conscious one associated  to the pattern signature; and
{\it b)\/}~an unconscious one which leads spontaneous movements constituting the drawing.
These two factors produce high variability, being the amount of signature variability 
strongly writer-dependent.
In fact, the signature {\it variability\/} or, conversely, the signature {\it stability\/} can be considered 
an important indicator for writer characterization\cite{Houmani2014}.
Houmani and Garcia-Salicetti\cite{Houmani2014} argue that signature stability is required in genuine signatures 
in order to characterize a writer: the less stable a signature is, the more likely it is that forgery 
will be dangerously close to genuine signatures for any classifier.
Also, complex enough signatures are required in order to guarantee a certain level of security, in the sense 
that the more complex a signature is, the more difficult it will be to forge it\cite{Houmani2014}.

Boul\'etreal and collaborators\cite{Bouletreau1998,Vincent2000} propose a signature complexity measure related 
to signature legibility and based on fractal dimension.
They classify writer styles into: highly cursive, very legible, separated,  badly formed and small 
writings, using only genuine signatures.
Unfortunately, such resulting categories were not confronted to classifiers for performance analysis.

\begin{figure}[hbt]
\centering
\includegraphics[width=.9\linewidth, angle=0]{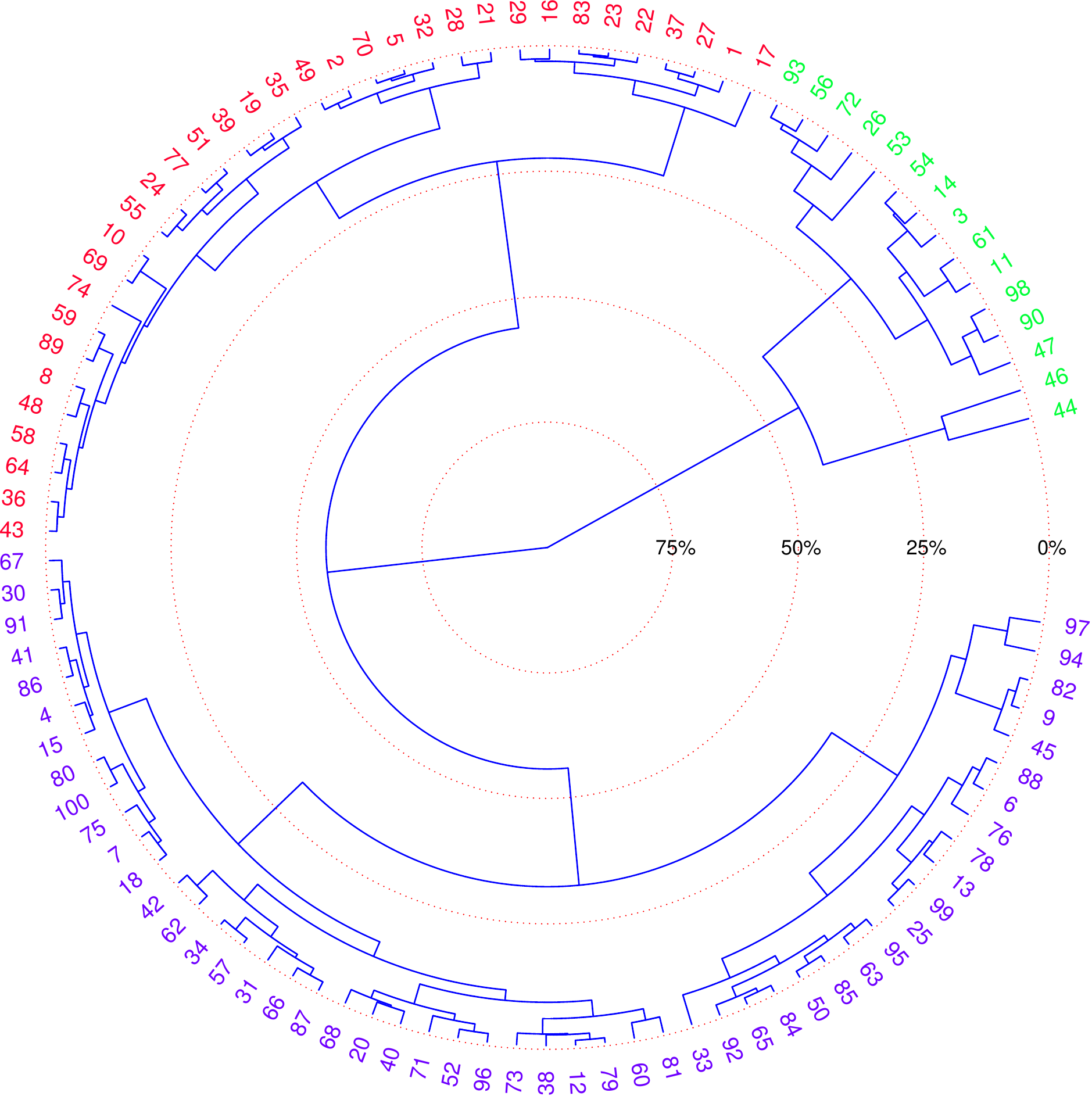}
\caption{
Neighbor-joining, rooted, circular dendrogram clustering of genuine signatures by Entropy: 
H1, H2, and H3, in red, blue, and green, respectively. 
}\label{Fig:DendroEntropy}
\end{figure}

We classify the genuine signatures based on causal Information Theory quantifiers: 
Normalized Permutation Shannon Entropy, 
Permutation Statistical Complexity and 
Permutation Fisher Information Measure of both $\mathbf X$ and $\mathbf Y$ trajectories
on each of the one hundred writers in the MCYT data base, and their $25$ original signatures.
The mean and standard deviation values were clustered using the neighbor-joining method and an automatic Hierarchical Clustering  
with the Euclidean distance-based dissimilarity matrix.
Each feature was treated independently, and the results are shown as circular dendrograms.
Figure~\ref{Fig:DendroEntropy} shows the results of clustering the Entropy.
We distinguish three classes of genuine signatures denoted by H1, H2, and H3. 

The H1 group is the first group to form, i.e., the one comprised of the most similar individuals.
It is formed below the $25\%$ level, and it is composed by two subgroups: H1A and H1B. 
The H1A group is formed exclusively by oversimplified signatures made by simple loops without identifiable letters. 
It encompasses the following subjects: 1, 16, 17, 22, 23, 27, 29, 37, 83. 
The same group is formed when the other features are used. 
The H1B group is comprised of the following subjects: 2, 5, 8, 10, 19, 21, 24, 28, 32, 35, 36, 39, 43, 48, 
49, 51, 55, 58, 59, 64, 69, 70, 74, 77, 89. 
Although these are simplified signatures, traces of letters and/or more complex curves appear and differentiate them 
from the members of the H1A group.

The H2 group is formed approximately at the $32\%$ level, and, again, it is comprised of two distinct groups: 
H2A and H2B. 
The subjects that make the H2A group are: 4, 7, 12, 15, 18, 20, 30, 31, 34, 38, 40, 41, 42, 52, 57, 60, 62, 
66, 67, 68, 71, 73, 75, 79, 80, 81, 86, 87, 91, 96, 100. 
It is composed by signatures with traces that resemble letters, but that are not perfectly identifiable, 
and that include circling traces of large or moderate size. 
Signatures in this group are kind of framed by large loops.
The H2B group is similar to the previous one, i.e., it is formed by signatures with large and medium size 
circling traces, but with more identifiable letters than in the previous groups. 
Names and surnames are more readable in this group than in previous ones. 
It is formed by the following signatures: 6, 9, 13, 25, 33, 45, 50, 63, 65, 76, 78, 82, 84, 85, 88, 92, 94, 95, 97, 99. 

The H3 group is formed at, approximately, the $43\%$ level by the fusion of two other highly unbalanced subgroups: 
one, H3A, with only two subjects (44, 46) and the other, H3B, with thirteen subjects (3, 11, 14, 26, 47, 53, 54, 56, 61, 
72, 90, 93, 98). 
These two clusters form at approximately the same level. 
The former is composed of calligraphic signatures where vertical traces predominate over horizontal ones. 
The latter is composed of highly cursive signatures, where separation between the surname and the family name 
predominates. 

The same results of clustering was obtained with the Manhattan (norm ${\cal L}_1$) and Maximum distances ($\mathcal L_\infty$ norm), showing that Entropy is an expressive and stable quantifier. 
Similar analyses were carried with the Permutation Statistical Complexity and Permutation Fisher Information 
(presented in figures Figs.~S5 and~S6 in the Supplementary Information).
Complexity produces the same clusters identified by Entropy, so it adds no new information.
The Fisher information measure forms the same H1A group that was identified by the Entropy, but with less 
cohesion, at about $15\%$. 
In other words, these nine subjects are more similar locally than globally. 
As with Entropy, three main groups form at similar levels. 
The members of these clusters are slight variations of those identified using Entropy, with very similar structure. 

Table~\ref{tab:tab-Measure-subject} presents 
the mean and standard deviation of the three quantifiers over the $25$ genuine and $25$ skilled forgery signatures 
(${\mathbf X}$ and ${\mathbf Y}$ time series) for each of the typical subjects, split in the three aforementioned types H1, H2 and~H3.
There are interesting tendencies in these data.
Genuine signatures present quantifiers values lower than those corresponding to forgery signatures, 
and the latter also exhibit larger standard deviation.
This could be explained by the imitative character of these signatures, however it deserves closer  studies.

\begin{sidewaystable}[hbt] \centering 
\centering
\begin{tabular}{@{\extracolsep{5pt}} ccccccc|cc|cc} 
\\ \cmidrule(r){6-11}
      &             &                    &  &  & \multicolumn{2}{c|}{Entropy}&\multicolumn{2}{c|}{Complexity}&\multicolumn{2}{c}{Fisher Information} \\ \midrule

$Type$ & $Sub-Type$ & Subject & Coordinate & Class & Mean & S.D. & Mean & S.D. & Mean & S.D. \\ 
\midrule 
\multirow{8}{*}{H1} & \multirow{4}{*}{H1A} & \multirow{4}{*}{22} & \multirow{2}{*}{${\mathbf X}$} &  F & $0.1568$ & $0.0052$ & $0.1490$ & $0.0039$ & $0.4688$ & $0.0070$  \\ 
		    &  			   &			 &                                &  G & $0.1519$ & $0.0019$ & $0.1457$ & $0.0015$ & $0.4766$ & $0.0035$  \\ \cmidrule(r){4-11}
		    &                      &  			 & \multirow{2}{*}{${\mathbf Y}$} &  F & $0.1595$ & $0.0071$ & $0.1511$ & $0.0052$ & $0.4665$ & $0.0097$  \\ 
		    &  			   &			 &                                &  G & $0.1512$ & $0.0042$ & $0.1447$ & $0.0037$ & $0.4734$ & $0.0046$  \\ 
\cmidrule(r){3-11} 
& \multirow{4}{*}{H1B} &\multirow{4}{*}{39} & \multirow{2}{*}{${\mathbf X}$} &  F & $0.2212$ & $0.0384$ & $0.1941$ & $0.0257$ & $0.4286$ & $0.0147$  \\ 
&                      &		    &			             &  G & $0.1749$ & $0.0037$ & $0.1620$ & $0.0028$ & $0.4497$ & $0.0029$  \\ \cmidrule(r){4-11}
&		       &		    & \multirow{2}{*}{${\mathbf Y}$} &  F & $0.2270$ & $0.0449$ & $0.1980$ & $0.0296$ & $0.4277$ & $0.0153$  \\ 
&		       &		    &			             &  G & $0.1776$ & $0.0043$ & $0.1644$ & $0.0031$ & $0.4491$ & $0.0035$  \\ 
\midrule 
\multirow{8}{*}{H2} & \multirow{4}{*}{H2A} &\multirow{4}{*}{60} & \multirow{2}{*}{${\mathbf X}$} &  F & $0.2482$ & $0.0593$ & $0.2112$ & $0.0365$ & $0.4212$ & $0.0107$  \\ 
		   & & &			&  G & $0.2010$ & $0.0056$ & $0.1803$ & $0.0040$ & $0.4331$ & $0.0031$  \\ \cmidrule(r){4-11}
		   & & & \multirow{2}{*}{${\mathbf Y}$} &  F & $0.2442$ & $0.0544$ & $0.2090$ & $0.0339$ & $0.4219$ & $0.0134$  \\ 
		   & & &			&  G & $0.2079$ & $0.0043$ & $0.1861$ & $0.0030$ & $0.4315$ & $0.0024$  \\ 
\cmidrule(r){3-11}
& \multirow{4}{*}{H2B}&\multirow{4}{*}{6} & \multirow{2}{*}{${\mathbf X}$} &  F & $0.2621$ & $0.0584$ & $0.2194$ & $0.0334$ & $0.4143$ & $0.0137$ \\ 
& &		   &			&  G & $0.2337$ & $0.0149$ & $0.2032$ & $0.0095$ & $0.4205$ & $0.0066$ \\ \cmidrule(r){4-11}
& &		   & \multirow{2}{*}{${\mathbf Y}$} &  F & $0.2648$ & $0.0538$ & $0.2218$ & $0.0304$ & $0.4136$ & $0.0134$ \\ 
& &		   &			&  G & $0.2314$ & $0.0102$ & $0.2018$ & $0.0067$ & $0.4211$ & $0.0050$ \\ 
 \midrule
\multirow{8}{*}{H3} &  \multirow{4}{*}{H3A} &\multirow{4}{*}{98} & \multirow{2}{*}{${\mathbf X}$} &  F & $0.3236$ & $0.0646$ & $0.2529$ & $0.0320$ & $0.3937$ & $0.0208$ \\ 
		 & &  &			&  G & $0.2707$ & $0.0101$ & $0.2268$ & $0.0064$ & $0.4106$ & $0.0032$ \\ \cmidrule(r){4-11}
		 &  & & \multirow{2}{*}{${\mathbf Y}$} &  F & $0.3204$ & $0.0794$ & $0.2497$ & $0.0388$ & $0.3970$ & $0.0208$ \\ 
		 &  & &			&  G & $0.2664$ & $0.0124$ & $0.2243$ & $0.0077$ & $0.4105$ & $0.0034$ \\ 
\cmidrule(r){3-11} 
& \multirow{4}{*}{H3B}&\multirow{4}{*}{46} & \multirow{2}{*}{${\mathbf X}$} &  F & $0.3514$ & $0.0641$ & $0.2691$ & $0.0294$ & $0.3940$ & $0.0156$  \\ 
& &		   &			&  G & $0.3480$ & $0.0282$ & $0.2720$ & $0.0156$ & $0.4019$ & $0.0047$  \\ \cmidrule(r){4-11}
& &		   & \multirow{2}{*}{${\mathbf Y}$} &  F & $0.3419$ & $0.0681$ & $0.2639$ & $0.0323$ & $0.3940$ & $0.0163$  \\ 
& &		   &			&  G & $0.3270$ & $0.0263$ & $0.2599$ & $0.0148$ & $0.4008$ & $0.0052$  \\ 
\bottomrule
\end{tabular} 
 \caption{Sample mean and standard deviation (S.D.) of the time series quantifiers for the 25 genuine (G) 
and 25 
skilled
forged (F) signatures, for each of the typical subjects: H1A, H1B, H2A, H2B, H3A, and H3B (same order as in Fig.~\ref{fig:MCYT-firmas}).} 
\label{tab:tab-Measure-subject} 
\end{sidewaystable} 

\begin{figure}[hbt]
\centering
\includegraphics[width=.9\linewidth, angle=0]{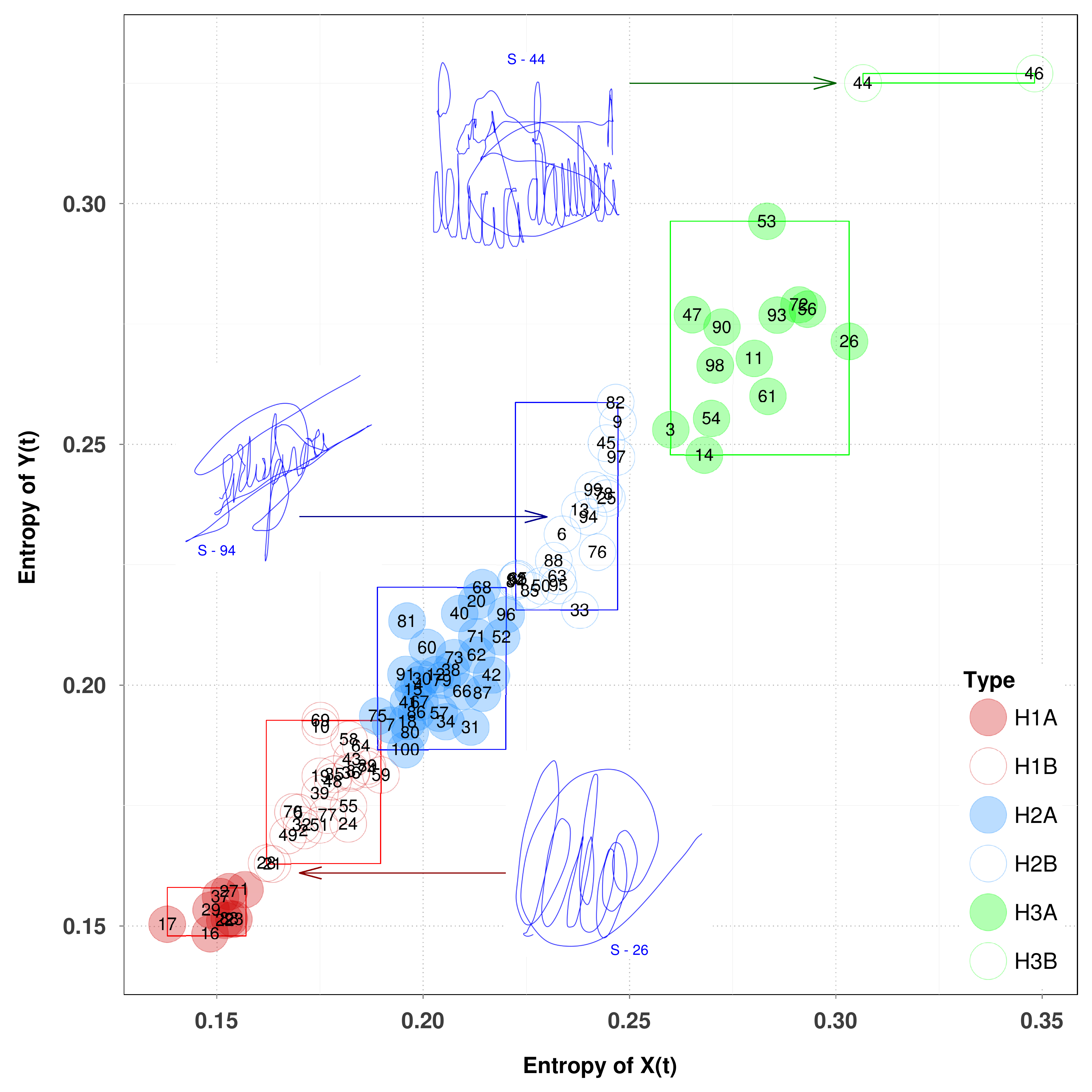}
\caption{Classification by the rule of the parallelepiped of genuine signatures using Entropy
(one signature example from each of the three groups is shown).
Each subject is identified by its ID.
}\label{Fig:EntropyClasificationBoxesEntropy}
\end{figure}

The classification into subclasses of genuine signatures was also carried by the parallelepiped 
algorithm\cite{Richards1999}, arguably the simplest model-free classification procedure.
Entropy leads to clusters with nice interpretability. 
Figure~\ref{Fig:EntropyClasificationBoxesEntropy} shows the regions that define the three classes 
identified by the dendrogram based on Entropy presented in Fig~\ref{Fig:DendroEntropy}. 
All subclasses are well separated by disjoint boxes, with the only exception of H1B and H2A that  
overlap slightly but without compromising the discrimination. 
The classes are preserved using this classification superimposed with Complexity and Fisher Information features; see Figs.~S7 and~S8 in the Supplementary Information.

\section*{Online signature verification}
\label{Sec-Verification}





The problem we have at hand consists in identifying suspicious signatures, given that we only have examples from genuine signatures.
This is due to the fact that, in practice, it is too expensive, too hard or even impossible to obtain a significant number of good quality forgery signatures for every possible individual in the data base.
This, thus, configures a One-Class classification problem.
Among the many ways of tackling such problems, Support Vector Machines (SVMs) are suitable for solving machine learning problems even in large dimensional feature spaces\cite{Campbell2011,Boser1992,Vapnik1995}. 

SVMs were introduced by Vapnik and co-workers\cite{Boser1992,Vapnik1998},
and extended by a number of other researchers.
Their remarkably robust performance with respect to sparse and noisy data makes them the choice in several applications.
A SVM is primarily a method that performs classification tasks by
constructing hyperplanes in a multidimensional space that separates cases of different class labels. 
SVMs perform both regression and classification tasks and can handle multiple continuous and categorical 
variables.
To construct an optimal hyperplane, a SVM employs an iterative training algorithm, which is used
to minimize an error function.

One-Class Support Vector Machines (OC-SVMs) are a natural extension of SVMs\cite{Scholkopf2001, Scholkopf2002}. 
The solution consists in estimating a distribution that encompasses most of the observations, and then labeling as ``suspicious'' those that lie far from it with respect to a suitable metric.
An OC-SVM solution is built estimating a probability distribution function which makes most of the observed data more likely than the rest, and a decision rule that separates these observation by the largest possible margin.
The computational complexity of the learning phase is intensive because the training of an OC-SVM involves a quadratic programming problem\cite{Boser1992}, but once the decision function is determined, it can be used to predict the class label of new test data effortlessly.

In our case, the observations are six-dimensional vectors: Entropy, Complexity and Fisher Information in each of the two directions, horizontal and vertical, and
we train the OC-SVM with genuine signatures.
Let ${\cal Z} = \{ z_1, z_2, \dots , z_N \}$ be the six-dimensional training examples of genuine signatures. 
Let $\Phi \colon {\cal Z} \rightarrow {\cal G}$ be a kernel map which transforms the
training examples to another space. 
Then, to separate the data set from the origin, one needs to solve the following quadratic programming problem:
\begin{equation}
\label{eq:oc-svm-1}
         \min_{{\mathbf{w}\in {\cal G},~\xi_i , b \in \mathbb{R}}} \qquad 
 \left\{\frac{1}{2} \|\mathbf{w}\|^2 + \frac{1}{\nu N} \sum_{i=1}^N \xi_i - b\right\}
\end{equation}
subject to         
\begin{align}
& \nu \in (0,1], \   \xi_i \ge 0, \  \forall i=1,\dots, N \label{eq:oc-svm-2}, \text{ and}\\
& (\mathbf{w} \cdot \Phi(z_i) ) \geq b- \xi_i, \  \forall i=1,\dots, N   ,     
\label{eq:oc-svm-3}
\end{align}
where $\xi_i$ are nonzero slack variables which allow the procedure to incur in errors.
The parameter $\nu$ characterizes the solution as
{\it a)\/} it sets an upper bound on the fraction of outliers (training examples regarded out-of-class) and,
{\it b)\/} it is a lower bound on the number of training examples used as Support Vectors.
We used $\nu=0.1$ in our proposal.

Using Lagrange techniques and a kernel function $K(z,z_i) = \Phi(z)^T \Phi(z_i)$, for the dot-product
calculations, the decision function $f(z)$ becomes:
\begin{equation}
\label{eq:oc-svm-primal-3}
f(z) = \text{sign}\left\{(\mathbf{w}\cdot \Phi(z)) - b \right\} = 
       \text{sign}\left\{ \sum_{i=1}^{N}~\alpha_i ~K(z,z_i) - b \right\}.
\end{equation}
This method thus creates a hyperplane characterized by  $\mathbf{w}$ and  $b$ which has maximal distance 
from the origin in the feature space $\cal G$ and separates all the data points from the origin.
Here $\alpha_i$ are the Lagrange multipliers; every $\alpha_i >0 $ is {\it weighted in\/} the decision function and thus ``supports" the machine; hence the name Support Vector Machine.
Since SVMs are considered to be sparse, there will be relatively few Lagrange multipliers with a nonzero value.

Our choice for the kernel is the Gaussian Radial Base function:
\begin{equation}
\label{eq:oc-svm-primal-4}
K(z_i,z_j) = \exp \Big(-\frac{1}{2\sigma^2} \|z_i - z_j\|^2
\Big) ,
\end{equation}
where $\sigma \in {\mathbb{R}}$ is a kernel parameter  and ${\|z_i - z_j\|^2}$ is the
dissimilarity measure; we used Euclidean distance.

The parameter $\sigma^2 = {10}$ was selected by 5-fold-cross validation, that its, the dataset is divided 
into five disjoint subsets, and the method is repeated five times. 
Each time, one of the subsets is used as the test set and the other four subsets are put together to form the training set. 
Then the average error across all trials is computed. 
Every observation belongs to a test set exactly once, and belongs to a training set four times.
Accuracy (ACC), Area Under the ROC Curve (AUC) and Equal Error Rate (EER) are used as performance measures~\cite{Kohavi1998}.

In the context of signature verification one-class classification problems, a false positive occurs when a genuine signature is erroneously classified as being atypical. 
The probability of false positive misclassification is the false positive rate, which is controlled by the parameters $\nu$ in the aforementioned OC-SVM formulation.
The parameter $\nu$ can be fixed a {\it priori} and it corresponds to the percentage of observations of the typical data which will be assigned as the Type~I Error. 

We used the LIBSVM (version 2.0) tool, linked with the R software, that supports vector classification and regression, including OC-SVM.\cite{Chang2001}
We used the standard parameters of the algorithm.

In order to assess the consistency of our procedure, and to promote the comparison with other methods reported in the literature, we evaluate the performance of the proposed verification system for different training samples: 
random samples of size $n$ ($n=5, 10, 14, 18, 22$)  of
genuine signatures were selected for each user.
Table~\ref{tab:ntrain} presents the average value of all performance metrics using $\sigma^2=10$.
ACC suggests that the larger the training sample, the better the performance is.
AUC presents a similar tendency, and its average is larger than $0.88$, indicating that our verification system produces an excellent classification. 

\begin{table}[hbt] 
\centering
\begin{tabular}{rccc} 
\toprule
$n$ & ACC ($\uparrow$) & AUC ($\uparrow$) &  EER ($\downarrow$)(\%)\\ 
\midrule
5  & 0.6940  & 0.8816 &  0.1890  \\ 
10 & 0.7678  & 0.8940 &  0.1711 \\ 
14 & 0.8144  & 0.8975 &  0.1634 \\ 
18 & 0.8250  & 0.8866 &  0.1731 \\ 
22 & 0.8389  & 0.8909 &  0.1632 \\ 
\bottomrule
\end{tabular} 
\vspace{0.25in}
\caption{Performance of the system trained with varying number $n$
of samples of genuine signatures.
$\uparrow$ and $\downarrow$ denote measures of quality (the higher the better) and of error (the smaller the better), respectively.
} 
\label{tab:ntrain}   
\end{table} 

As mentioned in the introduction, the two methodologies with best results are those based on Dynamic Time Warping (DTW) and Hidden Markov Models (HMM).
In the following we compare our proposal with these two recent state-of-the-art methods using the ERR(\%) over the same data base:
\begin{itemize}
\item Fierrez-Aguilar {\it et al.\/}\cite{Fierrez2005},  ERR(\%) = 2.12 (five training signatures; Global (Parzen WC)
and local (HMM) experts function);
\item Fierrez-Aguilar {\it et al.\/}\cite{Fierrez2007}, ERR(\%) = 0.74 (ten training signatures; HMM based algorithm);
\item Pascual-Gaspar  {\it et al.\/}\cite{Pascual2009}, ERR(\%) = 1.23 (five training signatures; DTW-bases algorithm,
result with scenario-dependent optimal features.
\end{itemize}
The results of our proposal using five (ten, respectively) training samples, are ERR(\%) = 0.19 (0.17, respectively).
Clearly, our system provides better performance using similar number of training signatures (see Table~\ref{tab:ntrain} for more details).

In the following we analyze the performance of the proposed procedure applied selectively to the pre-classified samples.
Table~\ref{tab:class} presents the performance of the system when applied to genuine pre-classified signatures. 
For all classes we observe that the larger the training sample, also the larger the average ACC is. 
The best average AUC are observed for the class H2, followed by H1 and H3.
This indicate that H2 signatures are easily identifiable. 
Note that the mean values of ERR(\%) for H2 are smaller than H1 and H3. 
The ERR(\%) values in H3 indicate that identifying forgeries in this class is hard.

%
\begin{table}[hbt]  
\centering
\begin{tabular}{crccc} \toprule
 Class & $n$ & ACC ($\uparrow$) & AUC ($\uparrow$) & EER(\%) ($\downarrow$)  \\ 
 \midrule 
\multirow{5}{*}{H1} 
 & 5 & 0.6758 & 0.8692 &  0.1976  \\ 
 &10 & 0.7566 & 0.8828 &  0.1812  \\ 
 &14 & 0.8039 & 0.8857 &  0.1717  \\ 
 &18 & 0.8217 & 0.8894 &  0.1662  \\ 
 &22 & 0.8277 & 0.8788 &  0.1631  \\ 
\midrule
\multirow{5}{*}{H2} 
 & 5 & 0.7059 & 0.8945 & 0.1784  \\ 
 &10 & 0.7819 & 0.9079 & 0.1548  \\ 
 &14 & 0.8284 & 0.9096 & 0.1509  \\ 
 &18 & 0.8327 & 0.8900 & 0.1734  \\ 
 &22 & 0.8515 & 0.8996 & 0.1608  \\ 
  \midrule
\multirow{5}{*}{H3} 
 & 5 & 0.6948 & 0.8653 &  0.2053 \\ 
 &10 & 0.7450 & 0.8720 &  0.2036  \\ 
 &14 & 0.7907 & 0.8832 &  0.1874 \\ 
 &18 & 0.8062 & 0.8686 &  0.1874  \\ 
 &22 & 0.8214 & 0.8889 &  0.1716  \\ 
\bottomrule
\end{tabular} 
\caption{Performance of the classification of pre-classified samples varying the number $n$ of 
samples of genuine signatures used for training; same coding as in Tab.~\ref{tab:ntrain}.} 
  \label{tab:class}
\end{table} 

\section*{Conclusions}
\label{Sec:Conclusions}

We proposed a procedure for identifying skilled forgery online handwritten signatures using time causal Information  Theory quantifiers and One-Class Support Vector Machines.
This is a competitive proposal from the computational viewpoint as it uses only the signatures coordinates, and it produces better results than state-of-the-art techniques.
The technique also produces meaningful classification of the input data, as it is able to separate different types of signatures.
To the best of our knowledge, this is the first time Information Theory quantifiers have been used for this problem.

The central contribution is the use of the Bandt and Pompe (BP) PDF symbolization  which is invariant to a number of transformations of the input data.
In fact, the original time series are pre-processed only to facilitate the signal sampling, and this scaling has no effect on the BP PDFs.
This representation, which is sensitive to the time causality, is able to capture essential dynamical characteristics of the signatures that lead to excellent discrimination between skilled forgery and genuine online handwritten signatures, despite the high variability the data possess.
Additionally, obtaining the BP PDFs is computationally simple and efficient.

Only six Information Theory features are required for the classification, three from each horizontal and vertical direction:
Shannon Entropy, Statistical Complexity and Fisher Information.
This contrasts many state-of-the-art works that require features in high-dimensional spaces, e.g. forty or even more.
As said, our proposal does not require highly specialized hardware able to capture signature speed, pressure, orientation etc.

The classification was performed by a One-Class Support Vector Machine trained with genuine signatures.
The learned rule is consistent with respect to the number of training samples, and with as few as five examples it surpasses the performance of recent successful techniques.
We assessed the performance of our proposal using the same data base employed in the current literature, with also the same measures of quality and error.

\section*{Acknowledgments}

The authors are grateful to CONICET, CNPq and FACEPE for partial funding of this research.
The Biometrics Research Lab (ATVS), Universidad Aut\'onoma de Madrid, provided the MCYT-100 signature corpus employed in this work.

\section*{Authors Contributions}

OAR, RO and ACF conceived and designed the research. 
OAR performed the numerical data analysis. 
RO and ACF performed the statistical analysis. 
OAR and RO prepared figures. 
OAR and ACF wrote the manuscript. 
All authors reviewed and approved the manuscript

\section*{Competing interests}

The authors declare no competing financial interests.

\end{document}